\documentclass[a4paper]{jpconf}
\usepackage{graphicx}
\begin{document}
\title{Multiparticle angular correlations: a probe for the sQGP at RHIC}

\author{W. G. Holzmann, N. N. Ajitanand, J. M. Alexander, P. Chung, M. Issah, R. A. Lacey, A. Taranenko, A. Shevel}

\address{Department of Chemistry, State University of New York at Stony Brook, Stony Brook, NY 11794-3400, USA}

\ead{wholz@mail.chem.sunysb.edu}

\begin{abstract}
A novel decomposition technique is used to extract the centrality dependence of di-jet 
properties and yields from azimuthal correlation functions obtained in Au+Au collisions 
at $\sqrt{s_{_{\rm NN}}}$=200~GeV. The width of the near-side jet shows very little 
dependence on centrality. In contrast, the away-side jet indicates substantial broadening 
as well as hints for for a local minimum at $\Delta \phi=\pi$ for central and 
mid-central events. The yield of jet-pairs (per trigger particle) slowly increases with 
centrality for both the near- and away-side jets. These observed features are compatible 
with several recent theoretical predictions of possible modifications of di-jet 
fragmentation by a strongly interacting medium. Several new experimental approaches, 
including the study of flavor permutation and higher order multi-particle correlations, 
that might help to distinguish between different theoretical scenarios are discussed.
\end{abstract}.

\section{Introduction}

QCD calculations on the lattice indicate a transition from hadronic matter
to a deconfined phase of quarks and gluons (QGP) at extremely high temperature and energy
density~\cite{Karsch:2002,Fodor:2001pe}. Heavy ion collisions can deposit large amounts of energy in 
a small collision volume. Thus, they provide excellent pathways to energy densities and 
temperatures pertinent to such a phase transition. 
Indeed, several recent results from Au+Au collisions at the Relativistic Heavy Ion Collider (RHIC) indicate 
the creation of matter with energy density believed to be sufficiently high for a state 
of matter with non-hadronic degrees of freedom to be formed~\cite{Adcox:2004mh}. 
The dynamical evolution of this matter appears to be hydrodynamically driven, implying that the produced 
medium is strongly interacting~\cite{Lacey:2001va,Shuryak:2004cy,Arkadij_ND,Gyulassy_qgp,Heinz,Heinz2}. 
This matter has been observed to strongly suppress the number of high transverse momentum particles~\cite{ppg003,ppg014,star_supp,ppg023} and to quench the away-side jet in central 
Au+Au collisions~\cite{Adler:2002ct}. The suppression is commonly ascribed to radiative energy loss of
hard scattered partons traversing the high energy density matter prior to the formation of hadrons
~\cite{bjorken,appel1986,blaizot_mclerran86,wang2,gyulassy,wang}. 

	The influence of parton-medium interactions on jet properties such as the number of jet-associated 
partner particles per trigger hadron, and jet-topologies, remains an open question of great current interest. 
Consequently, much recent effort is targeted at understanding how jets couple to the strongly interacting 
medium produced in energetic Au+Au collision 
at RHIC ~\cite{Ko,Hwa,Fries,Salgado,stoecker,stoecker2,shuryak,colorwake,Majumder:2004pt}. 
At the center of this discussion is the possible creation of jet induced ``conical flow" or ``mach shocks" 
analogous to a sonic boom in a fluid~\cite{stoecker,stoecker2,shuryak,colorwake}. 
If observed, such a phenomena could provide the means for reliable estimates of the speed of sound in 
the nuclear collision medium produced at RHIC.

	In this contribution, we utilize azimuthal angular correlation functions for charged hadrons to 
examine the influence of medium-effects on jet characteristics and yields. The use of correlation 
functions provides a novel and effective probe for such effects because they provide simultaneous 
access to both the (di-) jet signals and harmonic contributions arising from hydrodynamic evolution 
of the system. 

\section{Data Analysis}
The analysis presented here uses Au+Au data ($\sqrt{s_{NN}}$=200 GeV) 
provided by RHIC in the second running period (2001) and recorded by the PHENIX collaboration. 
The full PHENIX detector setup is described elsewhere~\cite{nim_1}. Charged tracks for this analysis 
were reconstructed in the two central arms of PHENIX, each of which 
covers 90 degrees in azimuth. Tracking was performed via a drift chamber and 
two layers of multi-wire proportional chambers with pad readout (PC1,PC3)\cite{nim_1}. 
A combinatorial Hough transform in the track bend plane was used for pattern 
recognition\cite{nim_2}. Most conversions, albedo and decays were
rejected by requiring a confirmation hit within a 2 $\sigma$ matching window in the PC3.
Collision centrality was determined using the beam-beam counters (BBC) and zero degree calorimeters (ZDC)
by cutting in the space of BBC versus ZDC analog response \cite{nim_3}.

Following the approach commonly exploited in HBT analyses', we define the azimuthal angular 
correlation function as the ratio of a foreground distribution $N_{cor}$, constructed 
with correlated particle pairs from the same event, and a background distribution $N_{mix}$, obtained 
by randomly pairing particles from different events within the same multiplicity and vertex classes. 
\begin{equation}
C\left(\Delta\phi\right) \propto \frac{N_{cor}\left( \Delta\phi
\right)} {N_{mix}\left( \Delta\phi \right)}.
\end{equation}
Here, $\Delta \phi=(\phi_{1}-\phi_{2})$ is the azimuthal angle difference between 
particle pairs formed with one hadron from a high-$p_{T}$ ``trigger" bin 
(hereafter labeled A) and another hadron from a lower $p_{T}$ selection (hereafter labeled B).
A particular asset of constructing a correlation function is that azimuthal acceptance 
and detector efficiencies automatically cancel in the ratio of foreground to background 
distribution. The area normalized correlation function then gives the probability 
distribution for detecting correlated particle pairs per event within 
the PHENIX pseudorapidity acceptance ($|\eta|<0.35$).

\begin{figure}[h]
\includegraphics[width=25pc]{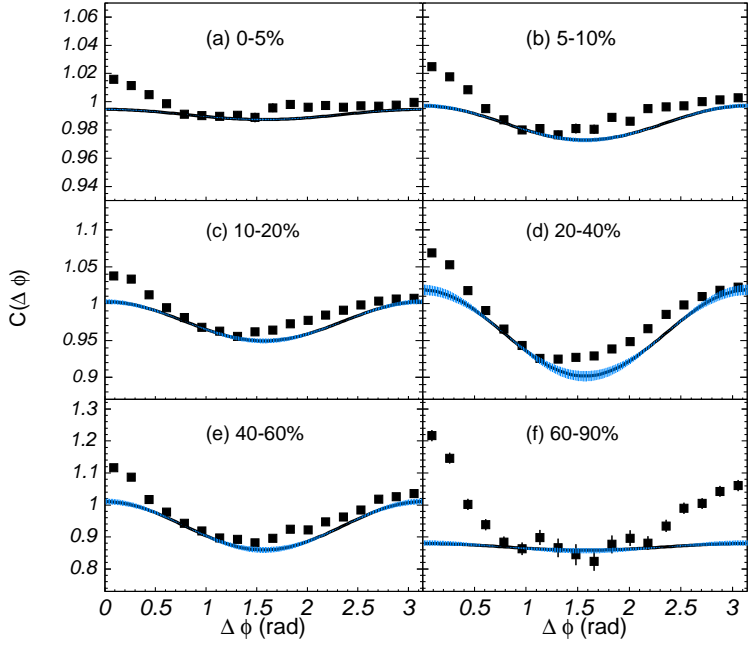}\hspace{2pc}
\begin{minipage}[b]{11pc}\caption{\label{Fi:fig01}Correlation functions, $C(\Delta\phi)$, for trigger particles in 2.5~GeV/c$<p_{T}^{A}<$4.0~GeV/c and associated particles in 1.0~GeV/c$<p_{T}^{B}<$2.5~GeV/c. The bands show the harmonic contribution within the systematic uncertainty. 
}
\end{minipage}
\end{figure}

Figure~\ref{Fi:fig01} shows the centrality dependence of di-hadron correlation 
functions obtained from Au+Au collisions with $p_T$ selections of $2.5 < p_T < 4.0$~GeV/c 
and $1.0 < p_T < 2.5$~GeV/c for trigger and associated hadrons respectively.  
The correlation function for the most peripheral event sample (cf. Fig.~\ref{Fi:fig01}f) 
shows two distinct peaks; a narrow peak at $\Delta \phi=0$ and a somewhat broader peak 
at $\Delta \phi=\pi$. This asymmetic pattern is compatible with the very well known 
azimuthal patterns expected for di-hadron correlations resulting from di-jet fragmentation.
In contrast to the peripheral events, the correlation functions for more central collisions 
show less prominent jet-driven asymmetries and appear to be dominated by large anisotropies
which can be linked to harmonic correlations resulting from elliptic flow.  
It is noteworthy that the correlation functions for central~-~mid-central collisions 
show a minimum below that expected for harmonic contributions ($\Delta \phi = \pi/2$). 
Such a shift strongly suggests an away-side jet which is significantly broader than in 
peripheral collisions. 

	A particularly important feature of the correlation functions shown in Figure~\ref{Fi:fig01} is 
the observation that they can be fully accounted for if one assumes that only two sources of correlations
contribute to the measured azimuthal correlation function. That is, a (di-)jet correlations and 
and harmonic correlations due to elliptic flow~\cite{Adler:2002ct,Ajitanand:2002qd,Chiu:2002ma,Adler:2002tq}. 
\section{Decomposition of jet and harmonic contributions}
It follows from the above discussion that a thorough investigation of possible 
modifications to the (di-) jet pair distribution in Au+Au collisions requires 
the separation of the jet-signal from the underlying harmonic ``background". 
Reliable procedures for such a decomposition are detailed in 
Refs.~\cite{Stankus_RP,Ajit_Methods}. Consequently, only the main steps of 
the procedure are outlined below.

It can be shown~\cite{Stankus_RP} that the pair correlations from the combination of 
flow and jet sources are given by 
\begin{equation}
C^{AB} (\Delta \phi ) = a_{o} [C^{AB}_H(\Delta \phi )] + J (\Delta \phi ), 
\label{eq1}
\end{equation}
where $C^{AB}_H(\Delta \phi )$ is a harmonic function of effective amplitude 
v$_{2}$, 
\begin{equation}
C_{H}^{AB} (\Delta \phi )_{ } = [1 + 2 v_{2} 
cos 2(\Delta \phi _{ }) ];  \  v_{2}=(v_2^A \times v_2^B).
\label{eq2}
\end{equation}
and J($\Delta \phi )$ is the (di-)jet function. 
By rearranging Eq.~\ref{eq1} one gets
\begin{equation}
J(\Delta \phi ) = C^{AB}(\Delta \phi ) - a_{o}C^{AB}_{H} (\Delta \phi ). 
\label{eq3}
\end{equation}
J($\Delta \phi$) can be extracted if the normalization (a$_{o}$) can be constrained 
and the harmonic amplitude of the background (v$_{2}$) can be reliably determined. 
To fix the value of a$_{o}$ we assume that the (di-)jet function has zero yield at 
the minimum (ZYAM) $\Delta \phi _{min}$.
%in the (di-)jet function. 
This is equivalent to demanding $a_0C_{H}^{AB}(\Delta{\phi}_{min}) = C^{AB}(\Delta \phi_{min})$.
However, no explicit or implicit assumption for the functional form of J($\Delta \phi$) is required. 
For the results presented here, the minimum is determined by requiring the background to coincide 
with a functional fit to the correlation function.
We obtain $v_2$ by measuring the single particle distributions relative to  
the reaction plane $\psi_R$. The reaction plane is reconstructed in the beam-beam counters
and corrected for dispersion following the procedures outlined in Ref.~\cite{Poskanzer:1998yz}.
An important aspect in this regard is the large (pseudo)rapidity separation ($\Delta \left|\eta \right|> 2.75$) 
between each BBC and the PHENIX central arms. This relatively large $\eta$ separation is believed 
to substantially reduce possible jet contributions to the determination of $v_2$~\cite{Adler:2003kt}.

Before applying the ZYAM decomposition technique to the analysis of data, we have tested its reliability
via a detailed set of Monte Carlo simulations that accounted for the $\phi$ and the $\eta$ acceptance of PHENIX~\cite{Ajit_Methods}.

After applying the ZYAM condition to the PHENIX data and utilizing the measured values of 
$v_2$ ($v_{2}=(v_2^A \times v_2^B$)), the harmonic
contribution to the correlation function is given by the solid bands in Fig.~\ref{Fi:fig01}.
The jet-pair distribution is then obtained by building the difference between the correlation function
and this harmonic contribution.
It can be shown, that the integral of this distribution is proportional to the fraction of jet-correlated particles
per event~\cite{Stankus_RP,Ajit_Methods}. This fraction can be extracted by building the ratio of the sum of $J(\Delta \phi)$ and the sum of $C(\Delta \phi)$ (over all bins in $\Delta \phi$),
\begin{equation}
PF = \frac{\sum_i{J(\Delta\phi_i)}}{\sum_i{C(\Delta\phi_i)}}
\label{Eq:JPF_decomp}
\end{equation}
One readily obtains the efficiency corrected pairs per trigger or conditional 
yield $CY$ from this fraction via multiplication by the ratio of the average 
number of detected particle pairs per event $\langle N^{AB}_{d} \rangle$, to the product 
of the detected singles rates $\langle N^{A}_{d}\rangle$, $\langle N^{B}_{d}\rangle $, 
followed by a final multiplication with the efficiency corrected singles rate 
$\langle N^{B}_{eff}\rangle$, for the lower $p_T$ bin B~\cite{Stankus_RP,Ajit_Methods}.
\begin{equation}
CY = PF \times \frac{\langle N^{AB}_{d} \rangle}{ \langle N^{A}_{d}\rangle \times \langle N^{B}_{d}\rangle}
\times \langle N^{B}_{eff}\rangle .
\label{Eq:CY_decomp}
\end{equation} 

\section{Results}

\begin{figure}[h]
\includegraphics[width=25pc]{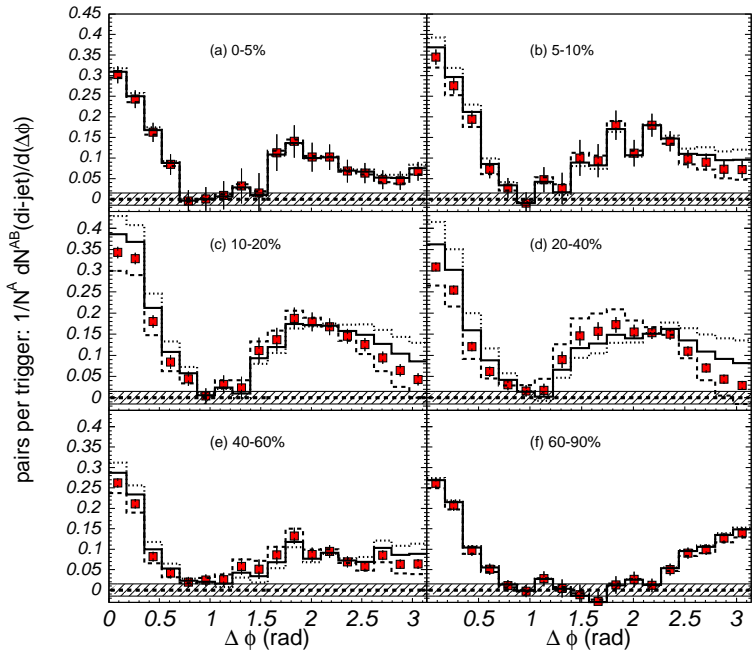}\hspace{2pc}
\begin{minipage}[b]{11pc}\caption{\label{Fi:fig02}ZYAM-subtracted jet-pair 
distributions $1/N_{trig}dN/d(\Delta\phi)$. The dashed(solid) histograms 
show the distributions that would result from subtracting the harmonic contribution after 
increasing(decreasing) the product $v_2 = v_{2}^{A} \times v_{2}^{B}$ by one interval of the systematic error.
The hatched area indicates the systematic error on the background determination (see text).}
\end{minipage}
\end{figure}

\begin{figure}[hbt!]
\begin{center}
\includegraphics[width=25pc]{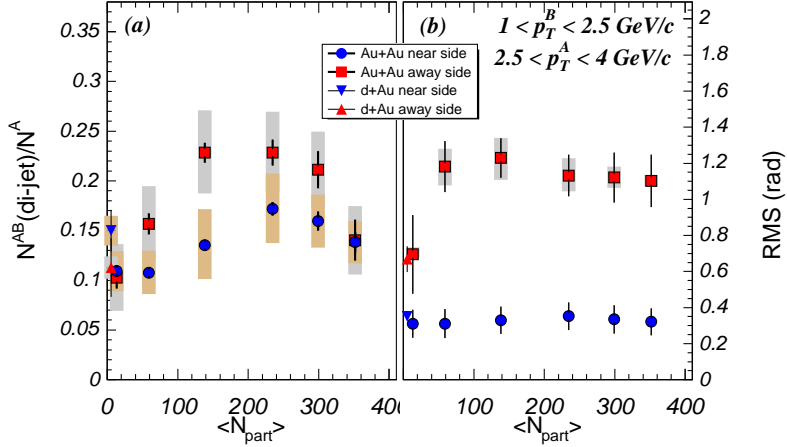}\hspace{2pc}
\caption[]{(a)~Per trigger yields for near- and 
away-side peaks in the jet pair distribution;
and (b)~Widths (RMS) of the peaks. The triangles
denote results for 0-20\% most central d+Au events from
a recent PHENIX analysis.\label{Fi:fig03}}
\end{center}
%\label{Fi:fig3}
\end{figure}

Figure~\ref{Fi:fig02} shows the centrality dependence of the extracted jet-pair distributions 
normalized to the number of trigger particles. The most peripheral bin exhibits an azimuthal pattern 
compatible with regular vacuum fragmentation of (di-)jets. That is, a near-side peak at $\Delta \phi=0$ 
and an away-side peak at $\Delta \phi=\pi$.
The jet-pair distributions obtained at other centralities reveal additional striking features. 
In order to quantify these features, we separate the jet-pair distributions into two regions
and assign $\Delta \phi=0 - \Delta \phi_{min}$ and $\Delta \phi_{min} - \Delta \phi=\pi$ to the 
near- and away-side jets respectively. Both pieces of the jet-pair distribution 
can then be characterized by their widths (RMS) and their respective per-trigger yields. 
The resulting values so obtained are plotted in Fig.~\ref{Fi:fig03} as a function of the 
number of participants. Similar results for the 0-20\% most central d+Au collisions are 
also included in Fig.~\ref{Fi:fig03} to facilitate a comparison between the Au+Au and 
d+Au systems.

	Let us first follow the centrality evolution of the near-side jet. 
The shape of this jet is essentially unchanged across centralities (cf. Fig.~\ref{Fi:fig02}). However it's per-trigger 
yield rises with increasing collision centrality suggesting some degree of modification 
[by the medium] which does not have a strong influence on the near-side jet shape. 
In contrast to the near-side jet, the shape of the away-side jet-pair distribution broadens 
significantly as one moves from peripheral events to mid-central events (cf. Fig.~\ref{Fi:fig02}). In addition, 
the mid-central distributions (20-40\% and 10-20\%) indicate an away-side jet distribution 
that exhibits a local maximum at $\Delta\phi=2\pi/3$ and an apparent minimum 
at $\Delta\phi=\pi$. These observations are consistent with a picture that involves little or 
no quenching of jets in peripheral events and strong centrality dependent medium modifications
of the away-side jet shape in the mid-central and central events. It is interesting to note 
that both the near- and away-side per trigger yields (cf. Fig.~\ref{Fi:fig03}) show a mild 
rise with increasing centrality, suggesting a possible medium induced modification of 
both near- and away-side jets. The prominent ``hump-backed" shape of the away-side jet is 
consistent with recent conjectures of a strong coupling between such jets and  
the high energy density matter that they traverse~\cite{stoecker,stoecker2,shuryak,Salgado}.

	It is instructive to study the influence of systematic variations in $v_2$ on the jet-pair
distribution. The solid (dashed) lines in Fig.\ref{Fi:fig02} indicate the conditional yield 
distributions that is extracted after subtraction of a $v_2$ product that is lowered (raised) 
by one interval of the systematic error respectively. The systematic error on $v_2$ is 
dominated by the uncertainty on the reaction plane dispersion. The dotted line shows the 
jet-pair distribution resulting from a subtraction of a $v_2$ product lowered by twice the 
systematic uncertainty. In the latter case, the local minimum
at $\Delta \phi=\pi$ is no longer significant. However, the away-side peaks for all 
centralities smaller than $60\%$, remain significantly broader than in the most 
peripheral event sample. 

\section{Summary of charged hadron results}

To summarize, we have used a novel decomposition technique to extract absolutely normalized jet pair 
distributions from correlation functions. Our results indicate jet pair distributions for 
peripheral Au+Au collisions which suggest very little, if any, medium induced
modification of jet properties. By contrast, the away-side jet measured in central and mid-central 
collisions show strong modifications resulting from interactions with the medium.
At this juncture, it is very interesting that the observed jet modifications are in qualitative 
agreement with several theoretical conjectures of how jet characteristics can be modified by a 
strongly interacting QCD liquid \cite{Salgado,stoecker,stoecker2,shuryak,colorwake}.

\section{Towards the Future }
	Future detailed studies of the modification of jet characteristic by the 
strongly interacting QCD medium will require new measurements as well as 
enhancements in our analysis techniques. In what follows, we give a few examples 
which show that such developments are already underway.
\subsection{Extraction of the jet function via harmonic extinction}

\begin{figure}[h]
\includegraphics[width=25pc]{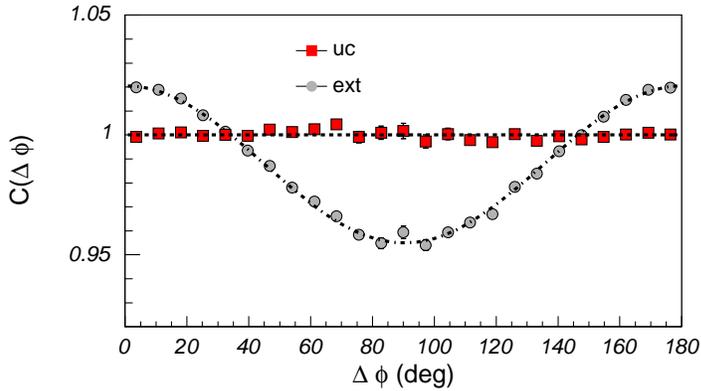}\hspace{2pc}
\begin{minipage}[b]{11pc}\caption{\label{fig5}Simulated correlation function for unconstrained particles (filled circles) and for 
 a trigger particle constrained within the cut angle $\phi _{c} = \phi _{xt}$ perpendicular 
 to the reaction plane (filled squares), see text. The results are for a pure harmonic 
 simulation with $v_2 \sim 0.16$.}
\end{minipage}
\end{figure}

For mid-central collisions, Fig.~\ref{Fi:fig01} shows pair correlations 
with strong harmonic components C$_{H}^{AB}(\Delta \phi )$ and relatively 
weak jet components J($\Delta \phi )$. Only after subtraction of C$_{H}^{AB}(\Delta \phi)$ is 
the true shape revealed for J($\Delta \phi )$ (cf. Fig.\ref{Fi:fig02} ). It would be desirable to 
select a data set that contained only those correlations due to the jets. 
In recent work we have developed a new technique which allows data selection such 
that the harmonic correlations are extinguished ((v$_{2})$ = 0) \cite{Ajit_Methods}.
This technique utilizes $\Delta \phi$ distributions where the trigger particles are
selected in a window perpendicular to the reaction plane~\cite{Voloshin:2004}. For these distributions $v_{2}$
can change phase and it is possible to select a cut angle $\phi _{c}$ so that $v_2$ = 0.
In Fig.~\ref{fig5} we demonstrate the extinguishing of harmonic correlations. 
The filled circles show the inclusive $\Delta \phi$ distribution obtained for a simulation
with an input $v_2 \sim 0.16$. The filled squares show the out-of-plane $\Delta \phi$ 
distribution after the cut angle $\phi _{c}$, is set to the extinction value, 
ie. $\phi _{c} = \phi _{xt}$ \cite{Ajit_Methods}. The latter distribution is flat 
and demonstrates that this technique gives a good method to determine the jet correlations 
directly from a data set, without the blurring effect from harmonic correlations 
mediated by the reaction plane. The latter technique provides clear advantages 
for future jet studies at RHIC. 
%
%Fig 5 :%%%
%
%
%
\subsection{Flavor permutations}
\begin{figure}[h]
\includegraphics[width=25pc]{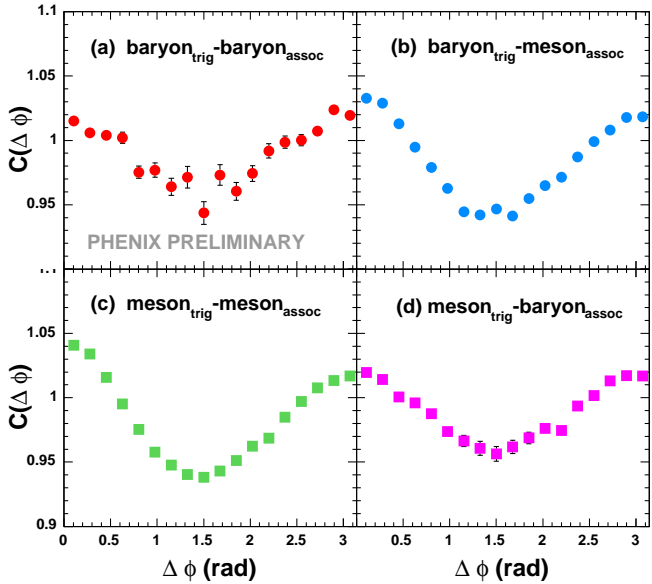}\hspace{2pc}
\begin{minipage}[b]{11pc}\caption{\label{Fi:fig05} Correlation functions for 
identified trigger particles ($1.5<p_{T}<2.5 GeV/c$) and identified partner 
particles ($0.8<p_{T}<1.5 GeV/c$). The centrality selection is 20-40\%.}
\end{minipage}
\end{figure}

The study of jet fragmentation and the influence of possible 
interactions of the jet with the medium in Au+Au collisions will undoubtedly require investigations 
which seek to map out the flavor dependence (ie. the dependence on particle species) of these effects. 
Correlation functions measured with fully identified particles at intermediate and high $p_T$ should 
serve as a powerful tool to unravel different aspects of the interaction between jets and the medium. 
Recently, we have begun to measure such correlation functions at RHIC.
Figure \ref{Fi:fig05} shows preliminary flavor permutation correlation functions for the centrality 
range 20-40\%. The trigger particle $p_T$ bin spans $1.5<p_{T}<2.5$ GeV/c and the assoicated 
particle has been selected from the range $0.8<p_{T}<1.5$ GeV/c. One can observe strong 
anisotropies and asymmetries [in these correlation functions] reminiscent of those discussed 
above for charged hadrons. More importantly, one can clearly see that these anisotropies and 
asymmetries strongly depend on the flavor permutation (ie. the the identity of each particle 
which is used in the pair correlation measurement). For example, the baryon-baryon correlation 
function shown in Fig.~\ref{Fi:fig05} (a) does not appear to exhibit any asymmetries
and shows a clear minimum at $\Delta \phi=\pi/2$, ie. the correlation function is essentially 
harmonic. This suggests little, if any, contributions from (di-) jet sources to the di-baryon 
correlation functions at these transverse momenta. On the other hand, the meson-meson 
correlation function shown in Fig.~\ref{Fi:fig05} (c) is relatively asymmetric, suggesting that 
jet fragmentation products are predominantly mesons. 
These data suggest that a detailed study of correlation functions as a function of flavor 
permutation holds much promise for providing detailed insights into jet fragmentation and 
its modification in a strongly interacting medium.

\subsection{Higher order multiparticle correlations}
In recent work we have also been pursuing analyses' which involve higher order multi-particle 
correlations. Such correlations have been shown to carry additional topological information compared
to a two particle correlation analysis \cite{LauretMulti}. Additional shape information from higher 
order correlations could potentially provide a powerful tool for distinguishing between different 
scenarios for jet-modification by the nuclear medium. For example, "mach cone" like structures 
would potentially manifest themselves in a rather distinctive pattern when correlating more than 
two particles. Such analyses for identified and unidentified particle multiplets are currently being 
pursued with vigor.

\end{document}